\documentclass[aps, twocolumn, reprint, longbibliography]{revtex4-1}

\usepackage{graphicx}% Include figure files
\usepackage{dcolumn}% Align table columns on decimal point
\usepackage{bm}% bold math

\newcommand{\dograph}[3]{
	\begin{figure}
	\includegraphics[#3]{#1}
	\caption{\label{fig:#1}#2}
	\end{figure}
}

\begin{document}

\title{Robust Fabry-Perot interference in dual-gated Bi$_2$Se$_3$ devices}
\author{A.D.K.~Finck$^{1}$, C.~Kurter$^{1,2}$, E.D.~Huemiller$^{1}$, Y.S.~Hor$^2$, D.J.~Van Harlingen$^1$}
\affiliation{$^1$Department of Physics and Materials Research Laboratory, University of Illinois at Urbana-Champaign, Urbana, Illinois 61801
\\
$^2$Department of Physics, Missouri University of Science and Technology, Rolla, MO 65409}
\date{\today}

\begin{abstract}
We study Fabry-Perot interference in hybrid devices, each consisting of a mesoscopic superconducting disk deposited on the surface of a three-dimensional topological insulator.  Such structures are hypothesized to contain protected zero modes known as Majorana fermions bound to vortices.  The interference manifests as periodic conductance oscillations of magnitude $\sim 0.1$ $e^2/h$.  These oscillations show no strong dependence on bulk carrier density or sample thickness, suggesting that they result from phase coherent transport in surface states.  However, the Fabry-Perot interference can be tuned by both top and back gates, implying strong electrostatic coupling between the top and bottom surfaces of topological insulator.
\end{abstract}
%\pacs{73.23.Ad, 73.63.-b, 74.45.+c}

\maketitle

%\section{Introduction}
The wave-like nature of electrons can manifest in mesoscopic devices through interference effects.  For example, if electrons traverse through a sufficiently clean sample between two leads with a finite probability of reflection at the sample-lead interface, then periodic conductance oscillations can occur as incident electrons interfere constructively or destructively with reflected carriers.  Such Fabry-Perot interference has been observed in carbon nanotubes \cite{Nature.411.665}, semiconducting nanowires \cite{NanoLett.10.3439}, and graphene \cite{Miao14092007}.  In all cases, special care during sample preparation must be taken to ensure quasi-ballistic transport.

Phase coherent transport in topological insulators (TIs) attracts attention for two reasons.  First, at the surface of a TI there are gapless states that behave as helical Dirac electrons \cite{RevModPhys.82.3045}, with the electron spin constrained to be perpendicular to momentum.  This spin-momentum locking generates a nontrivial Berry phase of $\pi$ as an electron moves in a complete circle around the Brillouin zone, which suppresses backscattering and protects the surface states from nonmagnetic disorder.  Second, if superconductivity is induced in the surface states of TIs, the resulting system will resemble a spinless $p_x + i p_y$ superconductor, with protected zero energy modes known as Majorana fermions bound to vortices \cite{PhysRevLett.100.096407}.  These Majorana fermions possess non-Abelian exchange statistics \cite{RevModPhys.80.1083, RevModPhys.87.137} and can be detected through interferometry \cite{PhysRevLett.102.216403, PhysRevLett.102.216404, PhysRevLett.103.237001}.  The complex geometry of the proposed interferometers as well as possible complications from bulk states encourage further studies of phase coherent transport in hybrid TI-superconductor devices.

Here, we report on transport studies of dual gated hybrid devices consisting of a superconducting disk deposited on the surface of a 3D TI.  We find clear signatures of gate-tuned Fabry-Perot oscillations.  The magnitude of the oscillations is essentially the same for lightly- and heavily-doped devices, implying that they originate from surface states of the TI rather than the bulk.  However, the oscillations can be tuned by applying a bias either to the top or back gate, suggesting the Fermi levels of the top and bottom surfaces are locked.  The resilience of the Fabry-Perot oscillations after multiple fabrication steps and over lengths of at least 800 nm as well as the option of using either top or back gates permit greater freedom in designing future TI interferometers.
%In the presence of a magnetic field applied perpendicular to the topological insulator surface, we find that the Fabry-Perot oscillations undergo a periodic phase shift of $\pi$.  This phase shift occurs whenever a magnetic flux quantum $\Phi_0 = h/2e$ is added to the region enclosed by the superconducting lead.  Unexpectedly, the phase shift is observed despite the lack of ferromagnetic regions surrounding the superconducting disk.  Our results suggest the presence of phase-coherent states encircling the superconductor, with implications for more advanced interferometric searches for Majorana fermions.

Single crystals of the 3D TI Bi$_2$Se$_3$ were grown by melting a mixture of pure Bi and Se in a stoichiometric ratio of 1.9975:3 (Bi:Se) in a vacuum quartz tube at 800 $^{\circ}$C.  Thin flakes (7-20 nm) of Bi$_2$Se$_3$ were exfoliated onto silicon substrates covered by a 300 nm thick SiO$_2$ layer.  Such thin flakes typically have a 2D carrier density of $N_{2D} \approx 10^{13} - 10^{14}$ cm$^{-2}$ and low temperature mobility $\mu \approx 10^2 - 10^3$ cm$^{2}$/V-s, as determined from measurements of separate Hall bar devices.  Order of magnitude variations in 2D carrier density of flakes of similar thicknesses suggest uneven Ca doping and thus variable bulk doping from sample to sample.  Weak anti-localization measurements of the Hall bar devices give typical phase-coherence lengths of $\ell_{\phi}= 300$ - 1000 nm at 10 mK.

Gold leads patterned by e-beam lithography were deposited through e-beam evaporation of 5 nm of Ti and 50 nm of Au.  Brief Ar ion milling is employed before metal deposition \emph{in situ} to ensure good contact between the Bi$_2$Se$_3$ and the leads.  A similar procedure was used to create low resistance ($< 50$ $\Omega$) contacts to Bi$_2$Se$_3$ that resulted in Josephson junctions with normal state resistances less than 100 $\Omega$ \cite{PhysRevB.90.014501, Kurter2013}.  The negligible contact resistance is confirmed when comparing two-terminal resistances to four-terminal resistance measurements in Hall bar devices.  A superconducting disk is inserted on the Bi$_2$Se$_3$ between the gold leads by DC sputtering of 50 nm of Nb at room temperature. The Si substrate acts as a back gate.  A top gate is created by covering the sample with 33 nm of Al$_2$O$_3$ via atomic layer deposition and evaporation of Ti/Au.  By measuring the change in carrier concentration of a separate Hall bar device with top gate bias, we find that the relative permittivity of the top gate dielectric is 8.75.

% Measurements of Au/Al$_2$O$_3$/Nb tunnel junctions reveal a superconducting gap of $\Delta = 1.5$ meV for our Nb films immediately after sputtering; in topological insulator-superconductor devices, the inverse proximity effect and additional nanofabrication processing will likely reduce the gap below this pristine value.

% Piece A: U2-III
% Piece B: U2-V
% Piece C: U2-IVB

% Piece D: U1-VI (good A-B oscillations)
% Piece E: U1-I (additional set of FP resonances)
% Piece F: U1-V (similar to VI)
% Piece G: U1-III (similar to I)

A typical device can be seen in Fig.~\ref{fig:overview}a and b.  The separation between the gold leads is 350 nm and the niobium disk is designed to have a diameter of 200 nm.  Here, the Bi$_2$Se$_3$ flake is 15 nm thick and 600 nm wide.  We studied two sets of devices: the first set (Piece A, B, and C) possessed a gold lead separation of 350 nm and Nb disk diameter of 200 nm.  The second set of devices (Piece D, E, F and G) were designed to have a lead separation of 800 nm and Nb disk diameter of 380 nm.  The samples in each set differ in conductance despite similar lateral dimensions, suggesting varying number of bulk states due to differences in thickness or variations in overall doping that were also observed in Hall bar devices.  Such discrepancies might be due to an uneven Ca doping.  In the first set, two of the devices (Piece A and Piece B) had higher resistance (a few k$\Omega$ at base temperature) while the third (Piece C) had a lower resistance (roughly 250 $\Omega$).  The second set of devices possessed resistances of 500 to 1000 $\Omega$. % Despite the order of magnitude difference of conductance, all devices showed similar magnitude of conductance variations from gate-tuned phase coherent transport.

\dograph{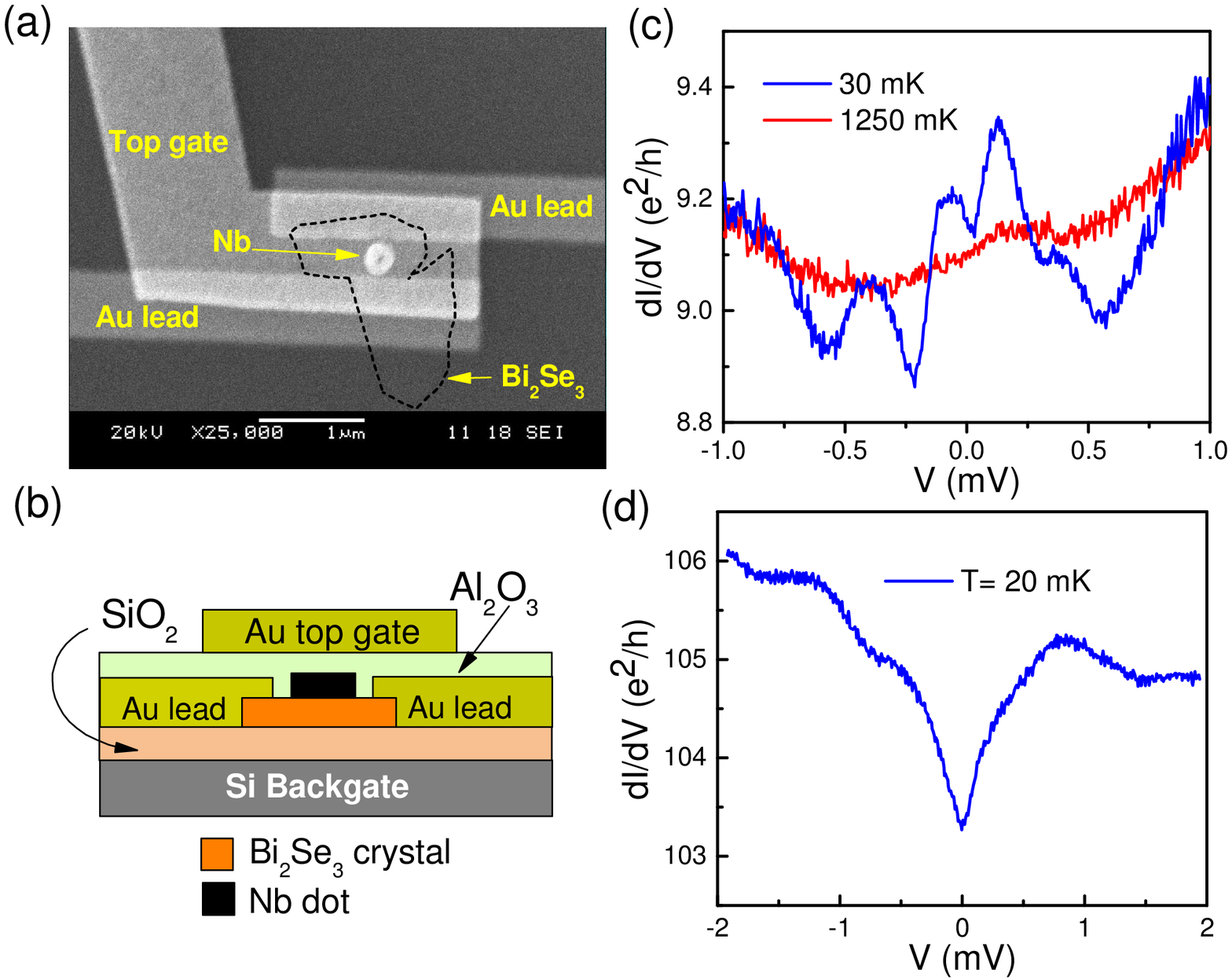}{(a) SEM micrograph of TI-superconductor interferometer.  Bi$_2$Se$_3$ flake (not visible) is outlined with dashed line.  (b) Cross-section of device, with Nb disk (black) located on top of Bi$_2$Se$_3$ flake (orange).  (c) Conductance vs source-drain bias $V$ for a high resistance device (Piece A) with no gate bias, at low and high temperature.  (d) Conductance trace for low resistance device (Piece C).}{width=3.25in, bb= 4 1 587 465}

%	U2-III / IVs vs Temperature 0 mT 7-18-2014
%	U2-V / IVs vs Temperature 0 mT 7-18-2014

The devices were thermally anchored to the mixing chamber of a cryogen-free dilution refrigerator with filtered wiring.  We perform transport measurements with standard lockin techniques.  Unless stated otherwise, all measurements were carried out at a base mixing chamber temperature of 20 mK.  As shown in Figs.~\ref{fig:overview}c and \ref{fig:overview}d, plots of conductance $dI/dV$ versus source-drain bias $V$ for Piece A and Piece C, respectively, reveal clear conductance resonances which become smeared out beyond $T=1$ K.  The peaks (dips) in conductance are signatures of Fabry-Perot resonances that occur due to constructive (destructive) interference between incident and reflected electrons in the TI.  For example, incident electrons can be reflected at the Bi$_2$Se$_3$-Au interface due to a mismatch of Fermi velocities for the two materials.  Restrictions on backscattering in the TI could be evaded either because of spin flips within the metallic leads or scattering between top and bottom surfaces.  Quantum interference originates from an additional phase that is accumulated by reflected electrons while traversing across the Bi$_2$Se$_3$ segment multiple times.  In the lowest order case of a single reflection at each Bi$_2$Se$_3$-Au (i.e. reflected electrons travel an additional distance of $2L$, where $L$ is the separation between the Au leads), this corresponds to a WKB phase of $2 k_F L$, where $k_F$ is the Fermi wave number.  This interpretation is confirmed by the gate-tuning of the resonances, as the Fermi energy (and thus Fermi wave number) evolves with gate bias.  In Fig.~\ref{fig:stability0mT}, we show a plot of conductance versus $V$ and top gate bias $V_{TG}$ from the high resistance sample Piece A, revealing the characteristic checkerboard pattern of Fabry-Perot resonances \cite{Nature.411.665, NanoLett.10.3439, Miao14092007}.  We note that while reports of Fabry-Perot oscillations typically have characteristic energy periods in the mV range, in our devices we are able to realize resonances with energy periods as small as 160 $\mu$V.

% stability0mT
\dograph{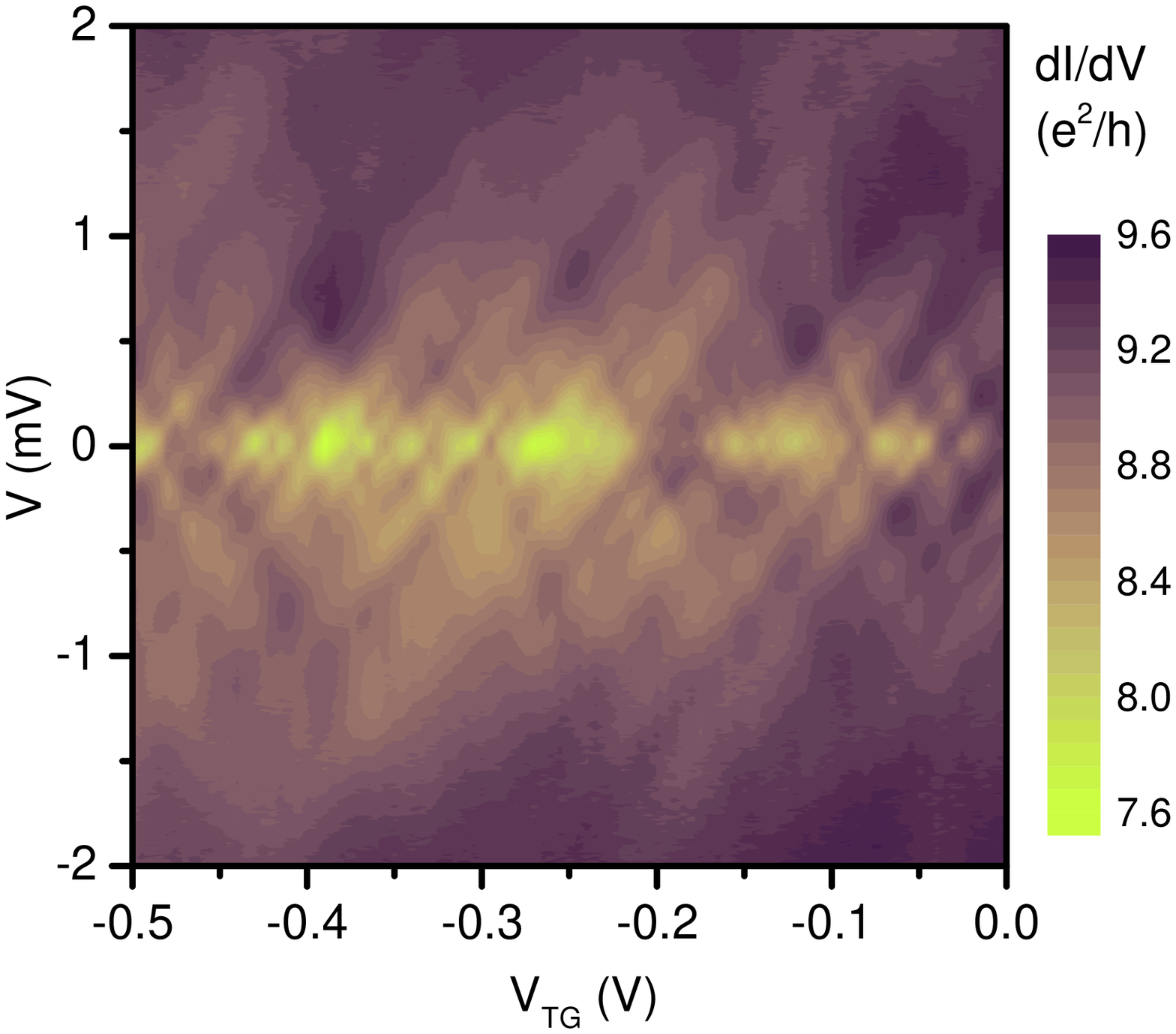}{Plot of $dI/dV$ vs source-drain bias $V$ and top gate bias $V_{TG}$ for Piece A.}{width=3.25in, bb= 17 18 502 444}

%	U2-III / IVs vs Top Gate 0 mT 7-6-2014
%	U2-V / IVs vs Top Gate 0 mT 7-5-2014
%	U2-IV-B / IVs vs Top Gate 0 mT 7-7-2014
The traces in Fig.~\ref{fig:stability0mT} show the sinusoidal behavior of Fabry-Perot resonances as the Fermi energy and wave vector change, superimposed on universal conductance fluctuations (UCFs) \cite{PhysRevLett.103.246601}.  At high source-drain bias (beyond the superconducting gap of niobium), the conductance oscillations have a characteristic peak-to-peak magnitude of $\approx 0.1$ $e^2/h$ and a top gate bias period of 32 mV.  Using the slope $\frac{dV}{dV_{TG}}$ of features in plots of $dI/dV$ vs $V$ and $V_{TG}$ to directly convert changes in top gate bias $\Delta V_{TG}$ into changes in Fermi energy via the equation $\Delta E_F = (\Delta V_{TG})(\frac{dV}{dV_{TG}})$, the conductance oscillations of Piece A have an energy periodicity of $\approx$ 0.34 mV.  Based on the predicted periodicity of $\Delta E = \frac{h v_F}{2 L}$ and $L \approx 350$ nm, this implies a Fermi velocity of $5.75 \times 10^4$ m/s, which is an order of magnitude smaller than the value extracted from angle-resolved photoemission spectroscopy of macroscopic pieces of Bi$_2$Se$_3$ \cite{Nat.Phys.6.960}.  The discrepancy could result from renormalization of the Fermi velocity due to coupling to bosonic modes such as phonons or surface plasmons \cite{PhysRevLett.110.217601}.  %We note that the observed energy period of the Fabry-Perot oscillations is smaller than for TI devices with smaller lead separation, but which lack superconducting disks in between the leads \cite{PhysRevX.4.041022}.  This helps to confirm that for the devices reported here, the quantum interference is determined by paths that travel completely from one gold lead to another, rather than being dominated by much shorter paths between one of the gold leads and the superconducting disk.

Similar to earlier results from hybrid TI-superconductor devices \cite{PhysRevX.4.041022}, we observe a bias-dependence of the oscillation period, with a frequency doubling occurring at a source-drain bias below the Nb superconducting gap.  This frequency doubling results from an interplay of phase coherent transport and multiple Andreev reflections in a ballistic system in contact with a superconductor \cite{PhysLett.4.151, PhysRevLett.16.453}. Low energy electrons in the TI undergo Andreev reflection at the Bi$_2$Se$_3$-Nb interface, generating reflected holes that also traverse the Bi$_2$Se$_3$ segment.  These reflected holes do not directly interfere with incident electrons.  Instead, they reflect off the Bi$_2$Se$_3$-Au interface and impinge upon the Nb again to reflect as electrons through a second round of Andreev reflection.  Such reflected electrons can interfere with incident electrons, but the phase difference is doubled to $4 L k_F$ as compared to conventional Fabry-Perot oscillations because quantum interference involves reflected particles that have traveled four times across the Bi$_2$Se$_3$ segment rather than merely twice.  Interestingly, we see signs of another doubling of the frequency near zero source-drain bias (see Fig.~\ref{fig:stability0mT}).  This could reflect higher order reflection processes \cite{Miao14092007} that is expected in relativistic Dirac systems, where specular Andreev reflection is permitted \cite{PhysRevLett.97.067007, PhysRevB.82.115312}.

%tgsweeps_vs_vsd
%\dograph{tgsweeps_vs_vsd}{Conductance of Piece A vs top gate bias for a variety of source-drain biases.  Note the prominent UCFs occurring at zero bias.}

%	U2-III / TG Sweeps vs VSD 0 mT 7-10-2014

%\section{Dual Gating of Fabry-Perot Oscillations\label{sec:dualgating}}

We next turn to simultaneous top and back gating of the Fabry-Perot oscillations in order to study their origin.  Under typical conditions, Bi$_2$Se$_3$ is not an ideal TI due to the presence of conducting bulk states \cite{NaturePhys.5.398} and trivial surface states from band bending \cite{NatComm.1.128}, either of which can coexist with the topologically non-trivial Dirac surface states.  The existence of both top and bottom surfaces with potentially different transport properties is an additional complication.  Thus, identifying the individual types of carriers through transport measurements can be difficult.  For example, studies\cite{PhysRevB.84.073109, PhysRevLett.109.116804} on Bi$_2$Se$_3$ films of varying thicknesses have extracted separate transport properties of bulk and surface states.  Total depletion of bulk carriers through aggressive chemical or electrochemical doping have revealed an ambipolar field effect that was attributed to the Dirac electrons \cite{NatPhys.8.460}.  Although in principle quantum oscillations can reveal the nontrivial Berry phase of the topological surface states \cite{Science.329.821}, in reality the large Zeeman coupling of bulk states can complicate this interpretation \cite{PhysRevB.85.033301}.

As we previously reported\cite{PhysRevX.4.041022}, the Fabry-Perot oscillations have a peak-to-peak magnitude of $\approx 0.1$ $e^2/h$, independent of sample thickness or total conductance.  This appears to rule out the bulk carriers as the source of the quantum interference.  However, it is not clear if the resulting Fabry-Perot oscillations result from either topological or trivial surfaces or if they are purely within the top or bottom surface of the TI device.

%tgsweeps_vs_vbg
\dograph{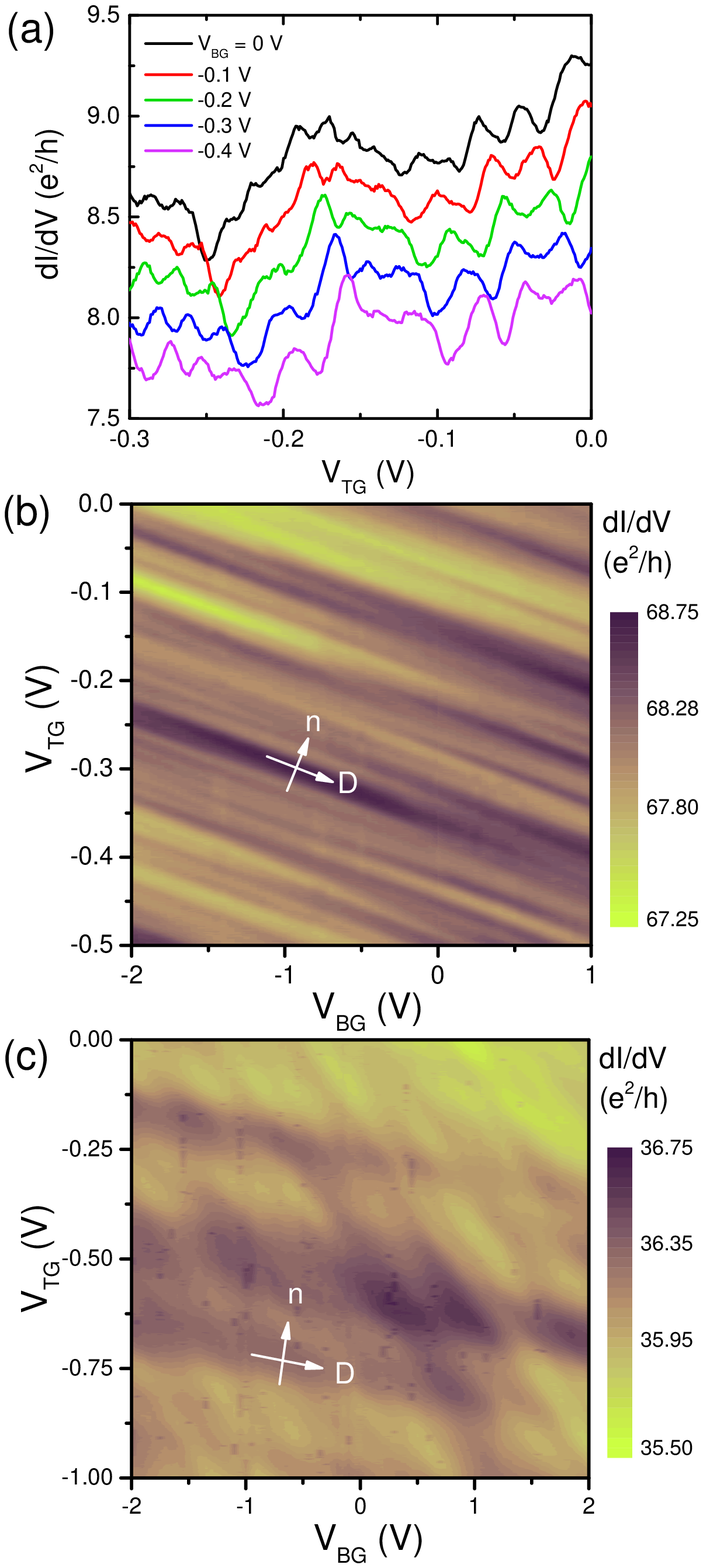}{(a) Conductance of Piece A vs top gate bias for a variety of back gate biases $V_{BG}$.  Curves are offset vertically for clarity.   (b) Plot of Piece D conductance vs top gate bias and back gate bias.  (c) Plot of Piece E conductance vs top gate bias and back gate bias.  Note the additional set of resonances with a different slope.  For (b) and (c), arrows indicate directions for either change in carrier density ($n$) or displacement field ($D$).}{width=3.25in, bb=3 6 346 774}

We consider applying biases to both gates.  One na\"ively expects that screening by the bulk carriers would cause the top gate to primarily influence the top surface while the back gate would mainly tune the Fermi energy of the bottom surface.  In Fig.~\ref{fig:tgsweeps_vs_vbg}a we show a series of top gate traces for different biases applied to the back gate from Piece A; similar results were obtained from all other interferometers.  We highlight two main observations from these data.  First, the positions of the resonances are linearly shifted by the back gate.  For Piece A, a 1 V bias applied to the back gate shifts the Fabry-Perot resonances by $\Delta V_{TG} = 78$ mV.  Similar shifts were observed for Piece B ($\Delta V_{TG} = 72$ mV per $\Delta V_{BG} = 1$ V) and Piece C ($\Delta V_{TG} = 62$ mV per $\Delta V_{BG} = 1$ V). Note that the ratio $\Delta V_{BG} / \Delta V_{TG} \approx 15$ for the relative efficiency of the top and back gates is close to the ratio of the gate capacitances $C_{top} / C_{bottom} \approx 20$.  Thus, the location of each Fabry-Perot resonance depends only on the total carrier density $n$, which can be tuned by either the top or back gate.  This implies that there is strong electrostatic coupling between the top and bottom surfaces \cite{NatPhys.8.460} due to imperfect screening of electric fields by surface or residual bulk carriers, such that applying a bias to the top gate will equivalently shift the Fermi level of the bottom surface and vice versa.  Here, it remains unclear whether the phase-coherent transport occurs in only one or both of the two surfaces.%We note that the shift in the Fabry-Perot resonances with back gate bias was also observed all devices.  %By using Fabry-Perot resonances to directly detect changes in the Fermi energy of the surface states induced by both top and bottom gates, we help to confirm that the Fermi energies of the top and bottom surfaces are locked\cite{NatPhys.8.460}.

A second observation from Fig.~\ref{fig:tgsweeps_vs_vbg}a is that certain features in the top gate traces can evolve in magnitude with back bias. An example is the peak that occurs near $V_{TG}=-0.2$ V at zero back gate bias.  As a negative voltage is applied to the back gate, this peak becomes more and more prominent.  Thus, the overall transport properties of Bi$_2$Se$_3$ is not purely a function of a single linear combination of $V_{TG}$ and $V_{BG}$, unlike the locations of the Fabry-Perot resonances.  Instead, such features vary with the displacement field, $D$, which is an antisymmetric function of $V_{TG}$ and $V_{BG}$ ($D = \alpha V_{TG} - \beta V_{ BG}$ for some positive constants $\alpha$ and $\beta$).  One possibility is that bulk carriers exhibit UCFs\cite{PhysRevLett.103.246601} that respond differently to top or back gating.

We further explore this second observation in Fig.~\ref{fig:tgsweeps_vs_vbg}b, where we plot conductance vs top gate bias and back gate bias for Piece D.  The arrows indicate the directions for either carrier density $n$ or displacement field $D$.  As with Piece A, the Fabry-Perot resonances can be tuned by either top or back gate, as reflected by the parallel bands of lines, which we assume to be along lines of constant $n$.  As hinted in Fig.~\ref{fig:tgsweeps_vs_vbg}a, we observe some features that also evolve with the displacement field.

We also note anomalous behavior in Piece E, which we show in Fig.~\ref{fig:tgsweeps_vs_vbg}c.  One set of the resonances are perpendicular to the $n$ axis, having the same slope as the resonances in Fig.~\ref{fig:tgsweeps_vs_vbg}b.  However, there is another set of resonances with a completely different slope.  This second set of resonances appears to be more strongly tuned by the back gate than by the top gate.  Such resonances might correspond to states that are partially screened from the top gate by either the metallic leads or other states within the TI.

We have presented a study of Fabry-Perot interferometry in hybrid TI-superconductor devices.  The magnitude of the Fabry-Perot resonances do not scale with bulk doping, suggesting that they originate from surface states.  The resonances can be tuned by both top and back gates, providing evidence of strong electrostatic coupling between the top and bottom surfaces.  Our results improve the understanding of phase coherent transport in TIs, which prove to be a useful system for studying quantum interference.
% We observe Fabry-Perot resonances with multiple gate frequencies, corresponding to higher order reflection processes, including those influenced by Andreev reflection at the TI-superconductor interface.
We are grateful for useful conversations with Liang Fu, Pouyan Ghaemi, Matthew Gilbert, and Shu-Ping Lee.  A.D.K.F., C.K., E.D.H., and D.J.V.H.~acknowledge funding by Microsoft Station Q.  Y.S.H.~acknowledges support from National Science Foundation grant DMR-12-55607.  Device fabrication was carried out in the MRL Central Facilities (partially supported by the DOE under DE-FG02-07ER46453 and DE-FG02-07ER46471).

\bibliography{topological}

\end{document}